\documentclass[a4paper,12pt]{iopart}
\usepackage{graphicx}
\usepackage{setstack,latexsym,amssymb}
\usepackage{amsfonts,amsthm}
\usepackage{epstopdf}
 
\newtheorem{theorem}{Theorem}

\newcommand{\us}{{\underline{ }}}   
\newcommand{\heun}{ {\rm HeunG}}   
\newcommand{\hyper}[2]{{}_{#1}F_{#2}}
   
\newcommand{\VGf}{\ensuremath{\mathcal{V}}}
\newcommand{\VGr}{\ensuremath{\overline{\mathcal{V}}}}

 \newcommand{\FGf}{\ensuremath{\mathcal{F}}}
\newcommand{\Hc}{\ensuremath{\mathcal{H}}}
\newcommand{\Rc}{\ensuremath{\mathcal{R}}}
\newcommand{\ddx}[1]{\frac{{\rm d}^{#1}}{ {\rm d}x^{#1}}}
 \newcommand{\mybinom}[2]{ \left(\!  {#1 \atop #2}\! \right) } 
  
\begin{document}

\title[Three friendly walkers]{Three friendly walkers\footnote{Dedicated to Tony Guttmann on the occasion of his 70th birthday. 
}}
\author{Iwan Jensen}
\address{School of Mathematics and Statistics,
The University of Melbourne, Vic. 3010, Australia}
\ead{ij@unimelb.edu.au}

\date{\today}                                           

\begin{abstract}
More than 15 years ago Guttmann and V\"oge  [J. Statist. Plann. Inference, {\bf 101}, 107 (2002)], 
introduced  a model of friendly walkers. Since then it has remained unsolved. 
In this paper we provide the exact solution to a closely allied model, originally introduced 
by Tsuchiya and Katori [J. Phys. Soc. Japan {\bf 67}, 1655 (1988)], which essentially 
only differs in the boundary conditions. 
The exact solution is expressed in terms of the reciprocal of the generating function
for vicious walkers which is a D-finite function. However, ratios of D-finite functions are inherently
not D-finite and in this case we prove that the friendly walkers generating function
is the solution to a non-linear differential equation with polynomial coefficients, it is in
other words D-algebraic. We then show via numerically exact calculations that the generating
function of the original model can also  be expressed as a D-finite function times the
reciprocal of the generating function for vicious walkers. We obtain an expression for 
this D-finite function in terms of  a $\hyper{2}{1}$ hypergeometric function with a rational pullback
and its first and second derivatives.
\end{abstract}



\noindent


\section{Introduction}

Consider $p$ directed walkers on the square lattice rotated through $45^{\circ}$ such that each walk take
 steps in the North-East direction $(1,1)$ or South-East direction $(1,-1)$. The walkers are labelled $k=1, 2, \cdots, p$.
 The positions of the walkers are given by the values of the ordinates $y$ after $t$ steps such that $y^k_{t}$ is the
 ordinate of the $k'$th walker after $t$ steps. The walkers are never allowed to cross but they may be allowed to share 
 vertices so $y^k_{t} \leq y^{k+1}_{t}$. We consider three versions of the walk  problem:
 
 \begin{enumerate}
 \item {\em Vicious walkers:} Walkers are not allowed to share a vertex and hence $y^k_{t} < y^{k+1}_{t}$.
 \item {\em Friendly walkers:} Two walkers may share vertices and edges for any number of steps. 
 \item {\em Super friendly walkers:} Any number of walkers may share vertices and edges for any number of steps.
 \end{enumerate}
 
\begin{figure}[h]
\begin{center}
\begin{picture}(450,130)
\put(0,0){\includegraphics[width=4.5cm]{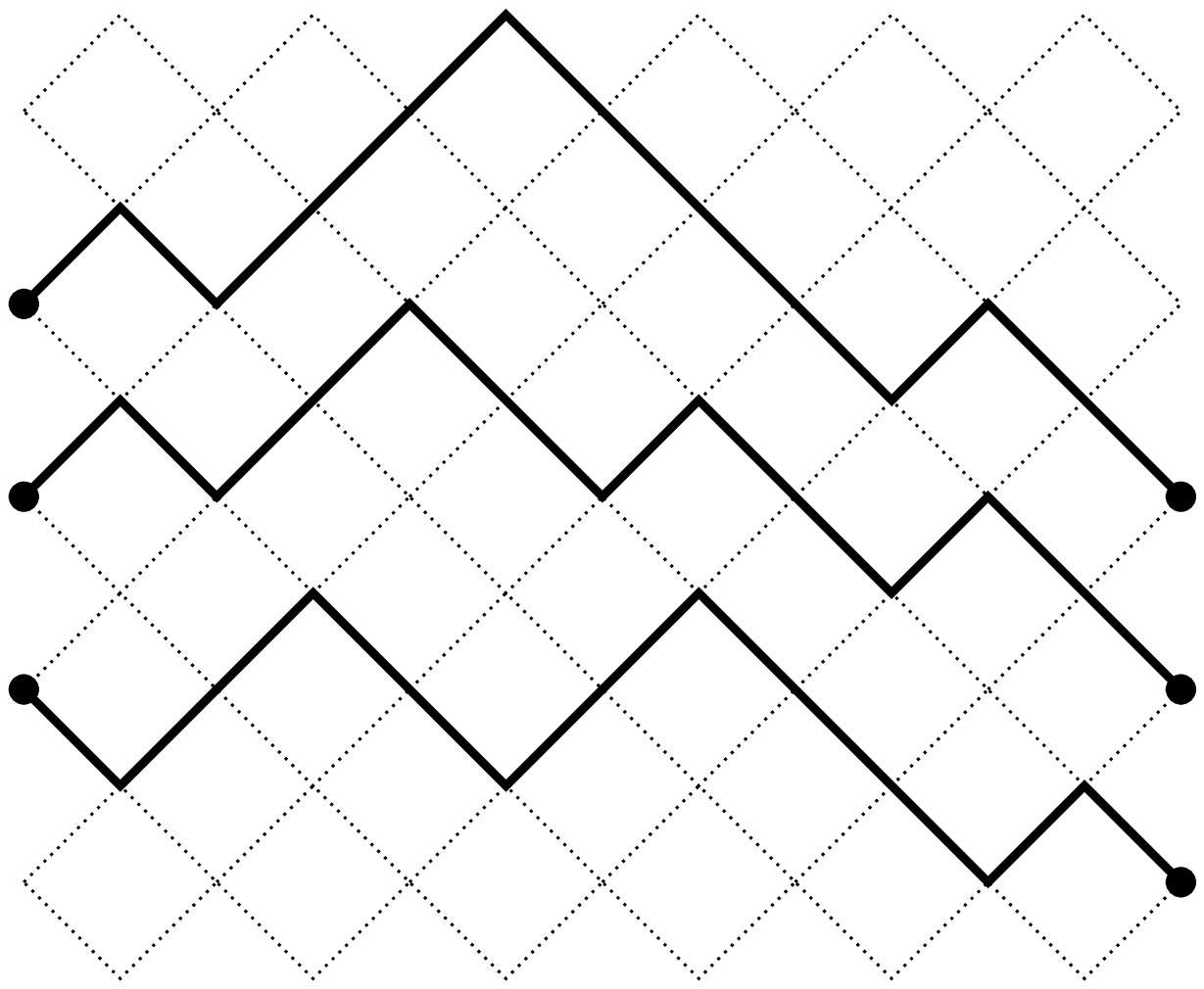}}
\put(150,0){\includegraphics[width=4.5cm]{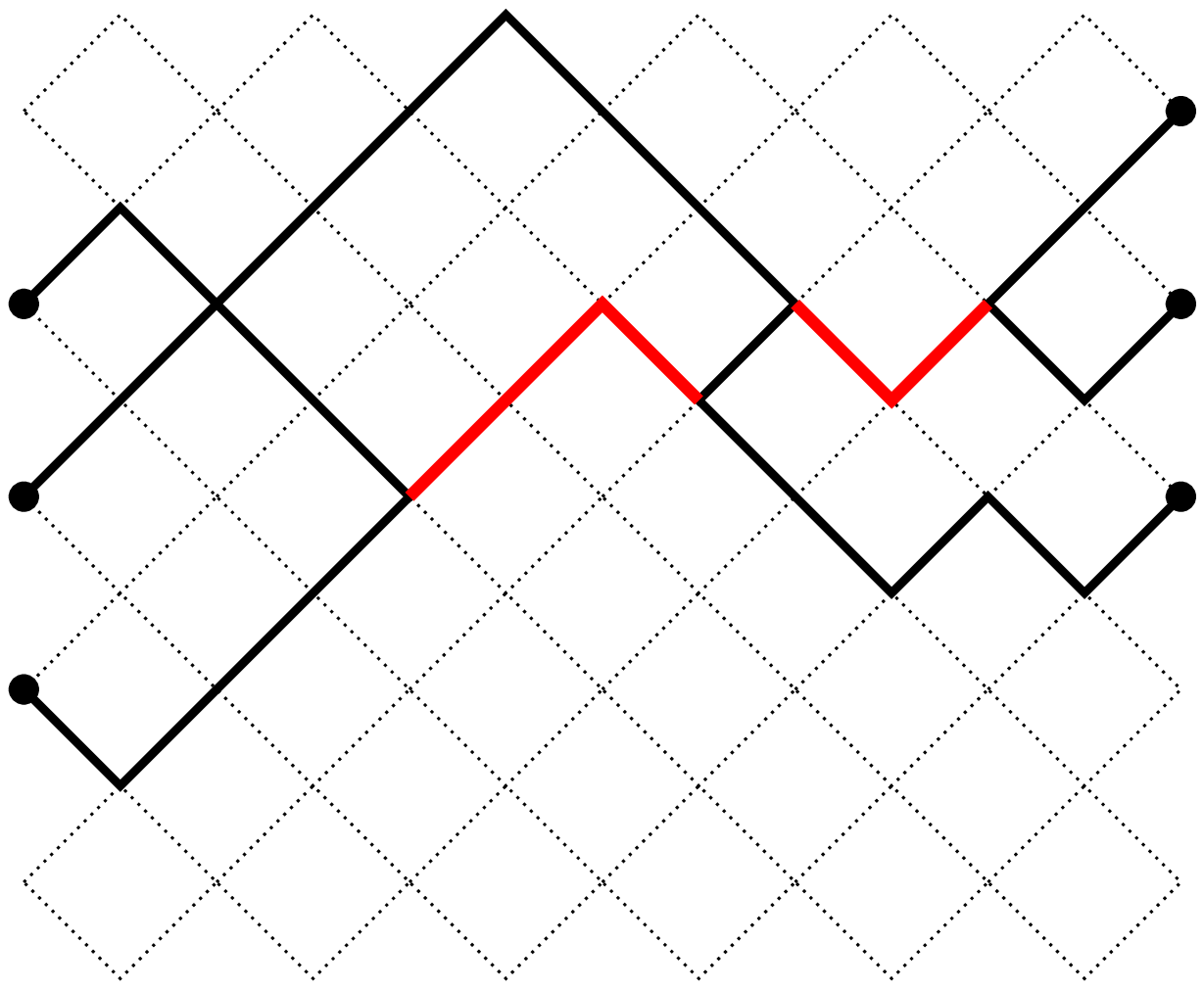}}
\put(300,0){\includegraphics[width=4.5cm]{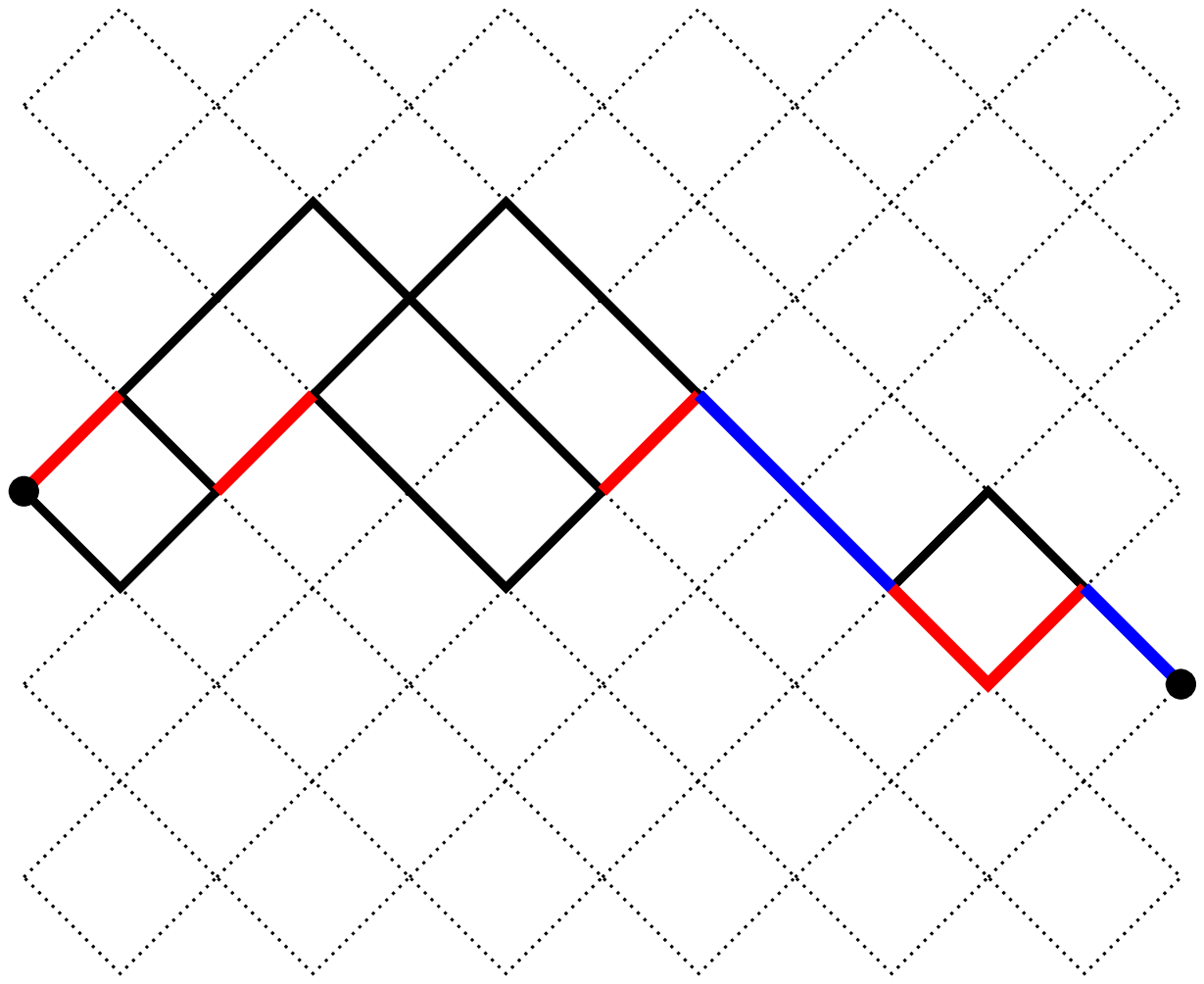}}
\end{picture}
\end{center}
\caption{\label{fig:walkex}
Examples of a vicious 3-watermelon, a friendly 3-watermelon and a super friendly 3-watermelon.
Black edges are singly occupied, red edges are shared by 2 walkers while blue  edges are shared by all 3 walkers.
}
\end{figure}

In the most general setting one can study walkers which start at a set of initial points  $y^k_{0}$ and end at a
set of end-points after $n$ steps $y^k_{n}$. However, in most cases one places some restrictions on these.
Typically one starts the walks at consecutive points such that $y^k_{0}=2(k-1)$. With no constraint
on the end-points one looks at so-called $p$-stars. If we force the walkers to terminate at consecutive
points we are looking at so-called $p$-watermelons. In this paper we study only watermelon configurations.
In the super friendly walker case it is perhaps more natural to start all walkers from the origin $y^k_{0}=0,\, \forall k$ and
also force them to end at the same vertex. Examples of the various models are given in figure~\ref{fig:walkex}.

Vicious walkers were introduced into the physics literature by Fisher \cite{Fisher1984} and the model has been extensively studied since.
Despite their simplicity directed walker models have intimate connections to
many profound and important physical and mathematical problems.
In physics  they are often used as simple lattice models of vesicles  and polymer networks 
\cite{Fisher1984,Fisher1991,Brak1992,Essam1995} and deep connections exist to lattice Green functions \cite{Guttmann1993a,Essam1993}.
The configurations of $p$ vicious walkers can be related to combinatorial objects such as plane partitions \cite{Gessel1989,Stembridge1990}, 
Young  tableaux \cite{Guttmann1998a,Krattenthaler2000,Krattenthaler2003} and symmetric functions \cite{Brenti1993}.  
Exact expressions for the number of configurations of $p$ vicious walkers of length $n$ have been obtained as simple  product formulae
in particular for the cases of stars and watermelons \cite{Essam1995,Guttmann1998a} and in some cases
exact closed form expressions have been obtained for the generating functions \cite{Essam1995}.
Friendly walkers were introduced by Guttmann and V\"oge \cite{Guttmann2002a} who named them $\infty$-friendly walkers.
The super friendly walker model was originally introduced by Tsuchiya and Katori in their studies of 
directed percolation \cite{Tsuchiya1998} and a version with interactions used to model polymer fusion or zipping transitions 
was solved exactly by Tabbara, Owczarek and Rechnitzer \cite{Tabbara2016}.  
If two walkers are allowed to share a vertex but not an edge one arrives at so-called osculating walkers 
which can be related to alternating sign matrices \cite{Brak1997}. 
An exact solution for the generating functions of stars and watermelons have been found  for $p=3$ \cite{BousquetMelou2006} 
and for general $p$ a constant term expression \cite{Brak2013} has been proved for the number of osculating
configurations of length $n$ with given starting and ending points.

In section~\ref{sec:3vw} we briefly review the results for vicious 3-watermelons and show that
the exact generating function obtained by Essam and Guttmann \cite{Essam1995} in terms
of a Heun function can in fact be expressed in terms of an $\hyper{2}{1}$ hypergeometric function
with a rational pullback and its derivative. In section \ref{sec:vf} we prove that the generating functions 
for a version of friendly 3-watermelons can be expressed in terms of the reciprocal of the generating
function of vicious 3-watermelons. We show that the friendly 3-watermelon generating function
is not D-finite but is in fact D-algebraic. In section~\ref{sec:numana} we provide results from
a numerical analysis of the singular behaviour of friendly 3-watermelons demonstrating that
their generating function have singularities of {\em infinite} order. Finally, in section~\ref{sec:GVsol} 
we report on numerically exact computations which show that the generating function of
the Guttmann-V\"oge model is the ratio of a D-finite function (the solution of a fifth order
inhomogenous ODE) and the  vicious 3-watermelon generating function. We show
that the numerator can be expressed in terms of the $\hyper{2}{1}$ hypergeometric function
appearing in the solution of vicious 3-watermelons and its first and second derivatives.

\section{Vicious 3-watermelons \label{sec:3vw}}

Essam and Guttmann \cite[Eq.~(63)]{Essam1995} proved that the generating function $\VGf_3(x)$  for vicious 3-watermelons 
is a solution to

\begin{equation}\label{eq:v3de}
\fl
x^2(1+x)(1-8x)G'' \;+\;x(8-42x-32x^2)G'\;+\;(12-40x-16x^2)G \;=\;12.
\end{equation}
which can be 
expressed in terms of  a Heun function\footnote{There appears to
be some minor misprints in the expression for the generating function in \cite[Eq.~(65)]{Essam1995}.} \cite{HeunBook}

{\renewcommand{\jot}{5mm}
\begin{eqnarray} \label{eq:v3}
\fl
\VGf_3(x) &\;=\;& \frac{1}{3x^3}\left[-1+x-3x^2+\heun\left(-\frac18,-\frac14; -1, -2, 2,-2;-x\right)\right] \nonumber \\
\fl
&\;=\;&  \frac{1}{3x^3}\left[-1+x-3x^2+\heun\left(-8,2; -1, -2, 2,-2;8x\right)\right]    \\
\fl
&\;=\;&  1+2\,x+6\,{x}^{2}+22\,{x}^{3}+92\,{x}^{4}+422\,{x}^{5}+2074\,{x}^{6}+
10754\,{x}^{7}  + \cdots, \nonumber
\end{eqnarray}}

\noindent
where we use the notation adopted in {\sc Maple}. $\VGf_3(x)$  has singularities at $x=x_c=1/8$ and $x=x_c=-1$ 
and at both  singularities the critical behaviour is of the form $(1-x/x_c)^3 \log (1-x/x_c)$. 
Assis \etal \cite{Assis2016a} found that  a $\heun$ function with integer coefficients 
could be recast in terms of an $\hyper{2}{1}$ hypergeometric function with an algebraic pullback. 
One of the authors\footnote{Jean-Marie Maillard in private e-mail exchange.}  has since 
told us that  generically  $\heun$ functions even with rational parameters do not correspond 
to series with integer coefficients nor can they be recast as series with integer coefficients.
Therefore if one sees a $\heun$ function whose series has integer coefficients it probably means that the   
$\heun$  function is not a generic $\heun$  with four singularities, it is in fact  a $\heun$   which
can be rewritten as a $\hyper{2}{1}$ with a pullback that wraps
the  four singularities of the  $\heun$  into the three singularities of the $\hyper{2}{1}$.
So we take a fresh look at the differential operator from (\ref{eq:v3de}) giving rise to the $\heun$ solution
\begin{equation}\label{eq:LHeun}
\fl \qquad
L_{\rm H} \;=\; x^2(1+x)(1-8x)D_x^2\;+\;x(8-42x-32x^2)D_x\;+\;(12-40x-16x^2).
\end{equation}
To check for  hypergeometric solutions we turn to the newly developed {\sc Maple} procedure 
{\tt hypergeometricsols} \cite{Imamoglu2015,Imamoglu2016}  which almost instantaneously  
finds that the solutions of $L_{\rm H}$ can indeed be expressed in terms of  
$\hyper{2}{1}$ hypergeometric functions. The solution corresponding to (\ref{eq:v3}) is

{\renewcommand{\jot}{8mm}
\begin{eqnarray} \label{eq:HeunF21}
\fl \qquad \heun\left(-8,2; -1, -2, 2,-2;8x\right) \quad = \nonumber\\
  \frac {\left( 1-8\,x \right)  \left( 1+x \right) ^{2}}{\left( 1-2\,x \right) ^{2}}  \;
\hyper{2}{1}\left(\left[\frac13,\frac23\right], \,[1],\,{\frac {27\, {x}^{2}}{ \left( 1-2\,x \right) ^{3}}}\right) + \nonumber \\ 
{\frac {x \left( 1-8\,x \right)  \left( 1+x \right) ^{2} \left(1+20\,x -8\,{x}^2 \right) }{ \left( 1-2\,x \right) ^{5}}\;
\hyper{2}{1}\left(\left[\frac43,\frac53\right],\,[2],\, {\frac {27\,{x}^{2}}{ \left( 1-2\,x \right) ^{3}}}\right)
}
\end{eqnarray}}
\noindent
Now the second $\hyper{2}{1}$ above is essentially the derivative of the first $\hyper{2}{1}$. In fact if we
let 
\begin{equation}\label{eq:HypSol}
\Hc(x) \;=\; \hyper{2}{1}\left(\left[\frac13,\frac23\right], \,[1],\,{\frac {27\, {x}^{2}}{ \left( 1-2\,x \right) ^{3}}}\right)
\end{equation}
and
\begin{equation}
\Rc(x)  \;=\;  \frac{\left( 1-8\,x \right)  \left( 1+x \right) ^{2}}{\left( 1-2\,x \right) ^{2}}
\end{equation}
then
\begin{equation}\fl \quad
\heun\left(-8,2; -1, -2, 2,-2;8x\right)   = \Rc(x) \Hc(x) - \frac{1}{24}(1-8x)(1-2x)^2 \Rc'(x) \Hc'(x).
\end{equation}

We shall see in  section~\ref{sec:GVsol} that the particular $\hyper{2}{1}$ hypergeometric function $\Hc(x) $
appears repeatedly in   3-watermelon problems and hence we shall often make use of the associated
differential operator $L_{\Hc}$ which annihilates $\Hc(x) $
\begin{eqnarray}\label{eq:HypDiff}
L_{\Hc} &\;=\;& x\left( 1+x \right)  \left( 1-8\,x \right) \left( 1-2\,x \right) ^{2}D_x^2 \;+ \nonumber \\
&&\left( 1-2\,x \right) \left( 1-12\,x -24\,{x}^{2} + 16\,{x}^{3}\right) D_x \; -\; 24\,x\left( 1+x \right)
\end{eqnarray}

It may  be of some interest to note that the second term in (\ref{eq:HeunF21}) can be re-written (simplified)
using Gauss's contiguous relations so that we get
{\renewcommand{\jot}{8mm}
\begin{eqnarray} \label{eq:HeunF21-2}
\fl \qquad \heun\left(-8,2; -1, -2, 2,-2;8x\right) \quad = \nonumber\\
  \frac {\left( 1-8\,x \right)  \left( 1+x \right) ^{2}}{\left( 1-2\,x \right) ^{2}}  \;
\hyper{2}{1}\left(\left[\frac13,\frac23\right], \,[1],\,{\frac {27\, {x}^{2}}{ \left( 1-2\,x \right) ^{3}}}\right) + \nonumber \\ 
{\frac {x   \left(1+20\,x -8\,{x}^2 \right) }{ \left( 1-2\,x \right) ^{2}}\;
\hyper{2}{1}\left(\left[\frac13,\frac23\right],\,[2],\, {\frac {27\,{x}^{2}}{ \left( 1-2\,x \right) ^{3}}}\right).
}
\end{eqnarray}}

It is also worth noting that $\Hc(x)$ can be replaced by the same $\hyper{2}{1}$ hypergeometric function 
but with a different rational pullback 
as a consequence of the   identity

\begin{equation}\label{eq:F21rel}
\fl \quad
\frac{1}{1-2x} \cdot \hyper{2}{1}\left(\left[\frac13, \frac23\right],[1], \frac{27x^2}{(1-2x)^3}\right)  \;=\;
  \frac{1}{1+4x} \cdot \hyper{2}{1}\left(\left[\frac13, \frac23\right],[1], \frac{27x}{(1+4x)^3}\right)
\end{equation}
where the two pullbacks $A(x)=27x^2/(1-2x)^3$  and  $B(x)=27x/(1+4x)^3$   are related by a
modular curve $\mathcal{C}= 0$ with
$$
\mathcal{C}\;=\; 8A^3B^3 -12A^2B^2(A+B) +3A\cdot B(2A^2+13A\cdot B+2B^2) -(A+B)(A^2+29A\cdot B+B^2) +27A \cdot B
$$
As usual   $z=0, 1,$ and $\infty$ are singularities of  the hypergeometric function $\hyper{2}{1}([a,b],[c],z)$, 
and we recall that the  hypergeometric differential equation has corresponding exponent pairs 
$\{0,1-c\}$, $\{0,c-a-b\}$, and $\{a,b\}$, respectively.  The condition that 
the two pullbacks equal 1, yield the singularities  $x= 1/8$ and $x= -1$. 
One may think that $x=1/2$ and $x=-1/4$  (such that the pullbacks $A(x)$ and $B(x)$ are  $\infty$)
are also singularities. This is not the case since $A(-1/4)=1/2$ while $B(1/2)=1/2$, i.e., where
one $\hyper{2}{1}$ appears to be singular the other clearly is not, 
and one  also sees that  the singular pre-factors must be cancelled by the $\hyper{2}{1}$.
Likewise, in (\ref{eq:HeunF21}) the singular pre-factors are cancelled  
when $x=1/2$  which isn't surprising since obviously $x=1/2$ is not a singularity of $\VGf_3(x)$. 

With this in mind one may ask if there is some way of re-writing $\Hc(x)$ and its companion in
(\ref{eq:HeunF21-2}) so the singular behaviour
becomes more transparent. One possibility is to use the Kummer relation
$$\hyper{2}{1}([a,b],[c],z)=(1-z)^{-b}\hyper{2}{1}([c-a,b],[c],z/(z-1)) $$
from  which we get
{\renewcommand{\jot}{8mm}
\begin{eqnarray} \label{eq:HeunF21-2}
\fl \qquad \heun\left(-8,2; -1, -2, 2,-2;8x\right) \quad = \nonumber\\
   (1-8x)^{1/3}(1+x)^{2/3}\hyper{2}{1}\left(\left[\frac23, \frac23\right],[1], \frac{-27x^2}{(1-8x)(1+x)^2}\right) + \nonumber \\ 
  {\frac{x \left(1+20\,x -8\,{x}^2 \right)} {(1-8x)^{2/3}(1+x)^{4/3} }\;
\hyper{2}{1}\left(\left[\frac23,\frac53\right],\,[2],\,\frac{-27x^2}{(1-8x)(1+x)^2}\right) 
}
\end{eqnarray}}

\noindent
Here at least we can clearly see that  $x= 1/8$ and $x= -1$ are singular. The integer values of the $c$
parameter means that the singularity at $\infty$ gives rise to an analytic solution and a solution with $\log z$.  

\section{Vicious and friendly walkers \label{sec:vf}}

In this section we consider  a variation of friendly 3-watermelons where the walkers start from the origin  
and end at the same vertex, but other than at the terminals there are never 3 walkers on the same vertex 
and never do 3 walkers share an edge.  We start by proving the following simple result.

\begin{figure}[h]
\begin{center}
\begin{picture}(350,130)
\put(0,0){\includegraphics[width=4.5cm]{ThreeVicious.eps}}
\put(160,60){$ \stackrel{\phi}{\longrightarrow}$}
\put(160,40){$ \stackrel{\displaystyle \longleftarrow}{\scriptstyle \phi^{-1}}$}
\put(200,0){\includegraphics[width=4.5cm]{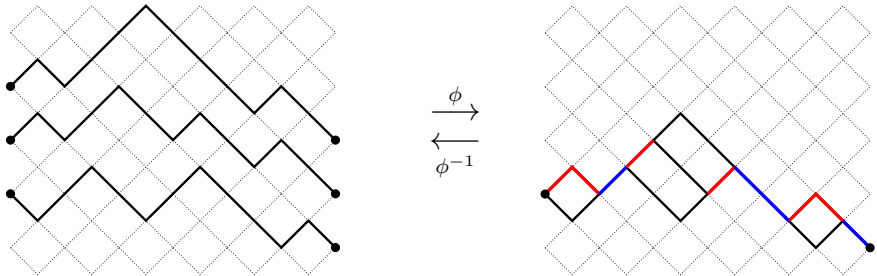}}
\end{picture}
\end{center}
\caption{\label{fig:vis2vf}
The mapping of a vicious 3-watermelon to a super friendly 3-watermelon and back. 
}
\end{figure}

\begin{theorem} \label{th:vicvf}
Vicious and super friendly $p$-watermelons are equinumerous.
\end{theorem}
\proof Let $S_V^n$ denote the (finite) set of vicious $p$-watermelons and $S_F^n$  the (finite) set of super friendly $p$-watermelons
of length $n$. Let $\phi$ be the function that acting on a vicious $p$-watermelon shifts the $k'$th walk downwards 
by $2(k-1)$ units., i.e, it maps the ordinates of a vicious walker $y^k_{t} \to  y^k_{t}-2(k-1)$ (see figure~\ref{fig:vis2vf}). 
Since for vicious walkers  $y^{k+1}_{t} - y^{k}_{t} \geq 2$ the new configuration is non-crossing and the walkers start at 
the origin and end at the same point. Hence it is a super friendly $p$-watermelon configuration. This shows that 
$\phi: S_V^n \to S_F^n$  and it is clearly injective. Conversely with the mapping $\phi^{-1}$  we take a friendly $p$-watermelon  and 
shift the $k'$th walk upwards by $2(k-1)$ units ($y^k_{t} \to  y^k_{t}+2(k-1)$) resulting in a vicious $p$-watermelon and again
this is an injective function. According to the Schr\"oder-Bernstein Theorem we have thus established a bijection between 
$S_V^n$ and $S_F^n$ proving that the two sets have the same cardinality.  
\qed

We can now proceed to derive an exact expression for the generating function $\FGf_3(x)$ for friendly 3-watermelons.

\begin{figure}[h]
\begin{center}
 \includegraphics[width=10cm]{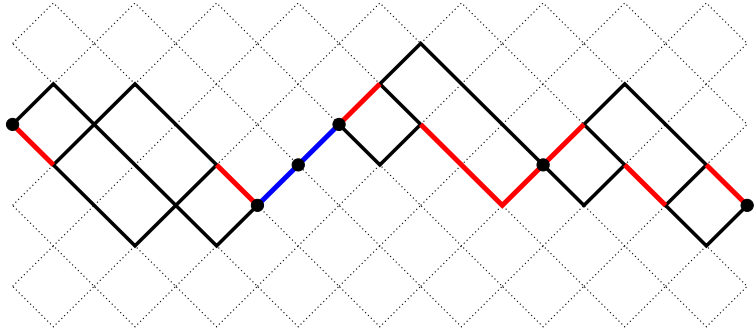}
 \end{center}
\caption{\label{fig:VFdec}
Decomposition of a super friendly 3-watermelon into 5 irreducible components $\omega_i$.  
}
\end{figure}

\begin{theorem} \label{th:3vfw}
The generating function $\FGf_3(x)$ for friendly 3-watermelons starting from the origin  and ending at the same vertex is

$$\FGf_3(x) \;=\; \frac{2(1-x)\VGf_3(x) -1}{\VGf_3(x)} \;=\;  2-2x -\frac{1}{\VGf_3(x)} $$
where $\VGf_3(x)$ is the generating function for vicious $3$-watermelons.
\end{theorem}
\proof By Theorem~\ref{th:vicvf} the generating function for super friendly 3-watermelons is $\VGf_3(x)$. Any
configuration of super friendly 3-watermelons can be decomposed into a sequence of irreducible components $\omega_i$
such that in each component the 3 walkers start at the origin and end on the same vertex   but never do the 3 walkers
otherwise share the same vertex (see figure~\ref{fig:VFdec}). Let $G(x)$ denote the generating function
for the the set of  irreducible components $\omega_i$. Since the walkers can take no steps we have

$$\VGf_3(x) = 1+ G(x)+ G(x)^2+\cdots = \frac{1}{1-G(x)},$$
which we invert to get
\begin{eqnarray*}
G(x) &=& \frac{\VGf_3(x)-1 }{\VGf_3(x)} 
\;\:=\;\: 2\,x+2\,{x}^{2}+6\,{x}^{3}+24\,{x}^{4}+110\,{x}^{5}+550\,{x}^{6}+\cdots.
\end{eqnarray*}
The term $2x$ comes from  three walkers simultaneously taking either North-East or South-East steps. These are not permitted 
friendly configurations so we remove these contributions. The remaining terms all arise from permitted configurations. Then adding
in the possibility of taking no steps we finally get
$$\FGf_3(x) \;=\; 1-2x+G(x) \;=\; \frac{2(1-x)\VGf_3(x) -1}{\VGf_3(x)}.$$~\qed

We can naturally also express $\FGf_3(x)$ in terms of a Heun function

\begin{equation}\label{eq:3fwheun}
\fl \qquad
\FGf_3(x) \;=\; \frac{2-4x+8x^2-3x^3-2(1-x)\heun(-8,2,-1,-2,2,-2,8x)}{1-x+3x^2-\heun(-8,2,-1,-2,2,-2,8x)}.
\end{equation}

Theorem~\ref{th:3vfw} immediately generalises to friendly  $p$-watermelons where up to $p-1$  walkers may share vertices and edges
for any number of steps.

\begin{theorem} \label{th:pvfw}
The generating function $\FGf_p(x)$ for friendly $p$-watermelons starting from the origin  and ending at the same vertex 
with up to $p-1$ walkers allowed to share vertices and edges for any number of steps is

$$\FGf_p(x) \;=\; \frac{2(1-x)\VGf_p(x) -1}{\VGf_p(x)} \;=\;  2-2x -\frac{1}{\VGf_p(x)}, $$
where $\VGf_p(x)$ is the generating function for vicious $p$-watermelons.
\end{theorem}
\proof Repeat mutatis mutandis the arguments of Theorem~\ref{th:3vfw}.~\qed

$\VGf_3(x)$ is a D-finite function. So $\FGf_3(x) $  is just a sum of a polynomial and the reciprocal of a  D-finite function, 
but $\FGf_3(x)$ is itself not D-finite. This is not unexpected since generically ratios of D-finite functions
are not D-finite, in fact in a quite remarkable paper Harris and  Sibuya \cite{Harris1985} proved that
if both $f$ and $1/f$ are D-finite then $f$ is algebraic. Now clearly given its logarithmic singular behaviour $\VGf_3(x)$ 
is not algebraic and hence  $\FGf_3(x)$ is not D-finite.  $\FGf_3(x)$ is however a solution of an 
algebraic differential equation, i.e., it is D-algebraic. Let $\VGr_3(x)=-1/\VGf_3(x)$ then using the {\tt Maple} package
{\tt GuessFunc} \cite{GuessFunc} developed by Jay Pantone one quickly finds that  $\VGr_3(x)$ is a solution to the 
non-linear D-algebraic equation
 
{\renewcommand{\jot}{5mm} 
\begin{eqnarray}\label{eq:vfda}
&&   x^2(1+x)(1-8x) \left[R \cdot R''-2 \left( R'\right) ^{2}\right] \;+\;  
 2x(4-21x-16x^2) R\cdot R' \nonumber \\
&& \;-\; \left( 12-40x-16{x}^{2} \right)R^2 \;+\;12\,  R^3 \;=\;0. 
\end{eqnarray}}
This result can be proven by making  the substitution $G(x)=-1/R(x)$  in the ODE (\ref{eq:v3de}) and expanding. 
Because of the second derivative there are terms  $1/R(x)^3$. 
Hence multiply the resulting equation (after the substitution) by $R(x)^3$, collect terms and the result  is (\ref{eq:vfda}).
Then, we find the expression for $\FGf_3(x) = 2(1-x) + G(x)$ by substituting $R(x) = G(x) - 2(1-x)$  in (\ref{eq:vfda})
and  evaluating derivatives. We thus prove that

\begin{theorem} \label{th:3vda}
The generating function $\FGf_3(x)$ for friendly $3$-watermelons starting from the origin  and ending at the same vertex 
 is a solution to the D-algebraic equation
 {\renewcommand{\jot}{5mm}
 \begin{eqnarray}
 \fl
&& x^2\left( 1+x \right)  \left( 1-8x \right)  F'' \cdot F 
 \,-\, 2\, x^2\left( 1-x^2 \right)  \left( 1-8x \right)    F''    \,-\, 2\,x^2 \left( 1+x \right)  \left( 1-8\,x \right) \left( F'\right)^2     \nonumber \\
 \fl
&&
+ \, 2\, x\left(4-21\,x-16\,{x}^{2} \right) F'\cdot F 
\,-\, 4\, x \left( 4-23\,x-9\,{x}^{2} \right) F'
\,-\, 12\, F^{3}  \\
\fl
&&
 +\, \left( 60-32x+16{x}^{2} \right)  F^{2}  \,-\, \left( 96-96x +132{x}^{2} \right) F
 \,+\, \left(48-64x +176{x}^{2}-48{x}^{3}\right) \;=\;0. \nonumber
  \end{eqnarray}}

 \end{theorem}

\section{Singular behaviour of $\FGf_3(x)$ \label{sec:numana}}

One can easily expand $\FGf_3(x)$ to thousands of terms and perform an asymptotic
analysis of the resulting power-series. Using biased differential approximants \cite{Jensen16a} we find compelling
evidence that  $\FGf_3(x)$ has a singularity at $x=x_c=1/8$ of {\em infinite} order with exponents that equal $3k, k\geq 1$,
so that the singular behaviour is

$$\sum_{k=1}^{\infty} (1-8x)^{3k} \left[\log (1-8x)\right]^{n_k},$$
where possibly $n_k=k$.  This is exactly the type of behaviour one would expect from
the expression (\ref{eq:3fwheun}) where barring some magic cancellations or other
simplifications one gets an infinite sum of powers of the $\heun$ function of (\ref{eq:v3}) which has
the singular behaviour $(1-8x)^3\log(1-8x)$. In table~\ref{tab:exp} we list as an example the exponent
estimates obtained from a single biased  differential approximant of order 16 with degrees of polynomials 
equal to 60 and biasing of order 8 at both $1/8$ and $-1$.  These results are quite remarkable in that
differential approximants (which essentially approximate a given function by a D-finite one) seems
very well-suited to extracting the critical behaviour of $\FGf_3(x)$ which, as we showed above, is in
fact not itself D-finite.

\begin{table}[h]
\caption{\label{tab:exp} Biased estimates for the leading critical exponents at the singularities $x_c=1/8$ and $x_c=-1$
as obtained from a single differential approximant of order 16 and degree 60 for friendly and $\infty$-friendly 3-watermelons.}
\begin{center}
\begin{tabular}{ll|ll}
\hline \hline
\multicolumn{2}{c|}{ $\FGf_3(x)$} & \multicolumn{2}{c}{ $\FGf_3^{\infty}(x)$} \\ \hline
$\quad x_c=1/8$  &  $ \quad x_c=-1$  & $\quad x_c=1/8$  &  $ \quad x_c=-1$   \\
\hline
$3+1.9\cdot 10^{-127}$  \qquad  &   $3+1.8\cdot 10^{-85} $  & $ 3-3.8\cdot 10^{-116}$ \qquad & $3-4.3\cdot 10^{-56}$\\
$6-1.3\cdot 10^{-126}$  & $4+9.4\cdot 10^{-20}$ & $6+2.8\cdot 10^{-92}$ &  $4-1.8\cdot 10^{-9}$ \\
 $9+2.7\cdot 10^{-109}$ &  $6+1.2\cdot 10^{-66}$  & $9+4.2\cdot 10^{-69}$ &  $6+6.9\cdot 10^{-38}$ \\
$12-5.0\cdot 10^{-83}$ & $7-2.0\cdot 10^{-12}$ & $12-2.4\cdot 10^{-47}$  & $6.99856$\\ 
$15+1.4\cdot 10^{-58}$& $9-1.6\cdot 10^{-45}$  & $15-6.3\cdot 10^{-28}$ & $9-3.8\cdot 10^{-22}$  \\
$18-1.6\cdot 10^{-36}$ & $10+3.0\cdot 10^{-4}$ & $18+3.8\cdot 10^{-12}$  & $12+4.8\cdot 10^{-9}$  \\
$21-4.0\cdot 10^{-17}$ & $12-4.9\cdot 10^{-27}$ & $21.012$ & $15.56563$\\
$24+9.0\cdot 10^{-5}$  & $15-1.9\cdot 10^{-11}$  & $58.275$ & $-0.75391$\\
\hline \hline
\end{tabular}
\end{center}
\end{table}

\section{Towards a solution for the Guttmann-V\"oge model \label{sec:GVsol}}

The model of $\infty$-friendly walkers introduced by Guttmann and V\"oge \cite{Guttmann2002a} is essentially 
identical to the model considered above except in boundary conditions. In the  $\infty$-friendly walker model
the walkers start and finish in a vicious configuration, that is $y^k_{0}=2(k-1)$ and if  $y^k_{t} -y^{k+1}_{t}=2\; (k=1,\cdots, p-1)$
then this is a valid $\infty$-friendly watermelon configuration of length $t$. 

The enumeration of these configurations is very fast since one has a polynomial time algorithm. 
One just keeps track of the ordinates  $y^k_{t}$. Clearly there is translational invariance in the ordinates so
one can always shift the ordinates so $y^0_{t}=0$ (alternatively it is the distances between consecutive 
walkers one needs not their actual positions). So with $p$ walkers and requiring a series to order $n$ one
needs on the order of $\mybinom{n/2}{p-1}$ configurations and hence for $p=3$ one has a polynomial time
algorithm of complexity $O(n^2)$. As one moves forward   each ordinate can change by $\pm 1$, i.e.,
$y^k_{t+1} =y^k_{t} \pm 1$ so that each configuration of ordinates at $t$ produces $2^p$ possible new configurations at $t+1$.
Any new configuration with $y^{k+1}_{t+1} < y^k_{t+1}$ is discarded since this would correspond to walkers crossing.
Since only two walkers may share a vertex we also discard configurations if  $y^{k+2}_{t+1} = y^k_{t+1}$. 
We  start in the `vicious' initial state $y^k_{0} =2(k-1)$ and   if $y^k_{t} -y^{k+1}_{t}=2\; (k=1,\cdots,p-1)$ 
we add the count of this configuration to the coefficient of $x^t$ in the generating function $\FGf_p^{\infty}(x)$.

So one readily calculates long series for the generating function $\FGf_3^{\infty}(x)$ for $\infty$-friendly 3-watermelons.
A series analysis shows singularities at $x_c=1/8$ and $x_c=-1$ and biased  differential approximants yields
a set of exponents equal to those for $\FGf_3(x)$ (see table~\ref{tab:exp}). So one may hope that  $\FGf_3^{\infty}(x)$ is also the
ratio of a D-finite function and $\VGf_3(x)$. Hence we form the function $H(x)=\FGf_3^{\infty}(x) \cdot \VGf_3(x)$
and lo and behold amazingly $H(x)$ is indeed D-finite being the solution of an inhomogeneous  linear ODE 
of order 5:

\begin{equation}\label{eq:def3v3}
\sum_{k=0}^5 P_k(x)\ddx{k} F(x) = P_{\rm I}(x), 
\end{equation}
where $P_5(x)=x^5 \left(1-8x\right)^3\left(1+x\right)^3 Q_{11}(x)$ with  $Q_{11}(x)$ a polynomial of degree
11 whose roots are  apparent singularities. The polynomials are listed in \ref{app:diffop}. 

The differential operator $L_5$ for the homogenous part has a direct sum decomposition into an order two and an order three
operator $L_5 = L_2 \oplus L_3$ as found using  the {\sc Maple} routine {\tt DFactorLCLM} from the {\tt DETools} package (the
operators are listed in \ref{app:L3L2}). The {\tt dsolve} routine finds that the operator $L_3$ has an exact solution in terms of a 
$\hyper{3}{2}$ hypergeometric function and two MeijerG functions. It turns out that the  MeijerG functions are not relevant solutions 
so we only list the hypergeometric solution

\begin{eqnarray}
S_1(x)&\;=\;& \frac{\left( 1+x \right) ^{9}}{x^9 \left( 1-8\,x \right) ^{3/2}} \cdot
\hyper{3}{2}\left(\left[\frac12,\frac32,\frac92\right]\!, \, \left[3,4\right], \,-64\,{\frac { x\left( 1+x \right) ^{3}}{ \left( 1-8\,x \right) ^{3}}}\right) 
\end{eqnarray}
{\tt dsolve} does not find a solution of $L_2$. The operator has singularities at $x_c=1/8$ and $-1$ with exponents 0 and 3 as
did $L_{\rm H}$. So we again turn to {\tt hypergeometricsols} which immediately finds that the relevant solution of $L_2$ can
be expressed in terms of  $\Hc(x)$

 \begin{equation}
\fl \qquad
S_2(x) \;=\; \frac { \left( 1-8\,x \right)  \left( 1+x \right)}{{x}^{10} \left( 1-2\,x \right) ^{2}} \left[x(1+x) P_1(x)\Hc(x) + 
 \frac{1}{12} (1-2\, x)P_2(x)   \Hc'(x) \right] 
 \end{equation}
with
\begin{eqnarray*}
P_1(x)&\;=\;&    137+595\,x-867\,{x}^{2}+1646\,{x}^{3}+298\,{x}^{4}-768\,{x}^{5}   \\
P_2(x)&\;=\;&     3+87\,x+3701\,{x}^{2}+7198\,{x}^{3}-5956\,{x}^{4}+   \\ 
&& 18962\,{x}^{5}-13544\,{x}^{6}+3248\,{x}^{7}-6144\,{x}^{8}   
\end{eqnarray*}
The particular solution to the  inhomogenous ODE is
$$S_{\rm P}(x) = \frac{1}{9x^9} (1+3x-6x^2+19x^3+6x^4+27x^5-27x^6)$$
We then have 
\begin{equation}
\FGf_3^{\infty}(x) \cdot \VGf_3(x) \;=\;   \frac19S_1(x) \;-\; \frac{1}{630}S_2(x)\;+\;S_{\rm P}(x).
\end{equation}

Next we take a closer look at the $\hyper{3}{2}$ solution to $L_3$. We first note that $L_3$ has singularities at 
$x_c=1/8$ (and $-1$) with exponents 0, 3 and 12 (9). So there we have that 0 and 3 combination again.
This could be a clue that the $\hyper{3}{2}$ is in fact expressible as a square of  $\Hc(x)$ and its derivatives.
To test this we turn to the {\tt DETools} package. The two routines we need are {\tt symmetric{\us}power} 
and {\tt Homomorphisms}.  The call  {\tt symmetric{\us}power}($L_{\Hc},2$) calculates a linear differential operator 
$M$ of minimal order which annihilates any product of solutions of $L_{\Hc}$, i.e., in particular  $\Hc(x)^2$ 
will be a solution of $M$. {\tt Homomorphisms}$(M_1,M_2)$ calculates (if one exists) a map $R$ (in general this will be a 
differential operator) such that $R$ maps the solutions of $M_1$ to those of $M_2$. Concretely this means that if $G(x)$ is a
solution of $M_1$, i.e., $M_1(G)=0$, then $R(G)$ is a solution of $M_2$, i.e., $M_2(R(G))=0$. The map $R$ 
is an  intertwiner between the two vector spaces of solutions of $M_1$ and $M_2$.  Indeed we find that the call
 {\tt Homomorphisms}({\tt symmetric{\us}power}$(L_{\Hc},2),L_3)$ calculates a second order differential operator 
 or intertwiner $I_3$, which shows that the solutions of $L_3$ can be expressed in terms of the solutions of
{\tt symmetric{\us}power}$(L_{\Hc},2)$. In particular we then get that the relevant solution of $L_3$ can
be expressed in terms of  $\Hc(x)^2$ and derivatives of  $\Hc(x)$. Concretely we can calculate the solution
with the call {\tt subs}($y(x)= \Hc(x)$, {\tt diffop2de}($I_3,y(x))$). We thus find with a bit of straightforward but tedious
calculation that

\begin{equation}\label{eq:S1F21}
6300\,S_1(x) = R_1(x)\Hc(x)^2+R_2(x)\Hc(x)\Hc'(x) +R_3(x)\left(\Hc(x)\Hc'(x)\right)'
\end{equation}
where the $R_k(x)$ are rational functions listed in \ref{app:S1F21}.

So at the end of all this we obtain an expression for $\FGf_3^{\infty}(x) \cdot \VGf_3(x)$ entirely
in terms of the simple hypergeometric function 
$$\Hc(x)=\hyper{2}{1}\left(\left[\frac13,\frac23\right], \,[1],\,{\frac {27\, {x}^{2}}{ \left( 1-2\,x \right) ^{3}}}\right)$$
and its first and second  derivatives.

\section{Conclusion, final remarks and outlook \label{sec:cum}}

In this paper we have found the exact solution to a friendly 3-watermelon problem. We proved that
the generating function $\FGf_3(x)$ can be expressed as the reciprocal of the vicious 3-watermelon 
generating function $\VGf_3(x)$ and showed that this result generalise to $p$ walkers. We then
showed that the generating function  $\FGf_3^{\infty}(x) $ for the Guttmann-V\"oge model of 
infinitely friendly 3-watermelons can be expressed as the ratio of a D-finite function and $\VGf_3(x)$ and
we obtained an exact expression for the numerator in terms  of a simple $\hyper{2}{1}$ hypergeometric function 
and its  first and second  derivatives.

We also had a look at the  Guttmann-V\"oge model of  2 friendly walkers \cite{Guttmann2002a} in which two walkers may
share an edge for a single step  after which they have to separate (similar to osculating walkers but on edges
rather than vertices). In this case our numerical analysis shows a critical behaviour very similar to that
of $\FGf_3(x)$  and   $\FGf_3^{\infty}(x)$, but as of yet we have not been able to find an expression
for the generating function in terms of  $\VGf_3(x)$. We hope to be able to do so in the future.

In future work we plan to study in some detail the  friendly $p$-watermelon problem and the problem
of $p$-stars as well. We hope that such studies can cast some light on the role of D-algebraic functions
in combinatorics and statistical physics.

Jay Pantone\footnote{Private communication.} has pointed out that it is possible to use the guessed D-finite equation 
for $H(x)= \FGf_3(x)\VGf_3(x)$ and the known D-finite equation for $\VGf_3(x)$ to recover a conjectured D-algebraic 
equation for $\FGf_3(x)$ by using the process of differential elimination. 
The discouraging thing is that the resulting equation is somewhat monstrous. It contains a total of 133 different
terms (involving products  of powers of $\FGf_3(x)$ and its derivatives) each with a polynomial coefficient of degree
up to 51 or so. So it would take between 7000 and 8000 series terms to guess the equation. The highest order
derivative occurring in the D-algebraic equation is of order 7 and triple products occur. There are terms such as
$\FGf_3^{(5)}\FGf_3^{(7)}$ or $\FGf_3^{(4)}\FGf_3^{(5)}\FGf_3^{(6)}$, where $\FGf_3^{(k)}$ is the $k$'th derivative
of  $\FGf_3(x)$. Naturally, one would hope that a simpler D-algebraic equation can be found but we have not  
been successful as yet.

\ack
The author would like to thank Jay Pantone for granting access to {\tt GuessFunc} prior to official release,
advice on how to use it and in particular for pointing out how to derive the D-algebraic equation (\ref{eq:vfda});
Jean-Marie Maillard for realising that $\FGf_3^{\infty}(x) \cdot \VGf_3(x)$ is D-finite and invaluable help 
in analysing the corresponding ODE.
The author was supported by the Australian Research Council via the discovery grant DP140101110.

\appendix{

\section{The differential operator $L_5$ \label{app:diffop}}

The polynomials $P_k(x)\; (k=0,\ldots,5)$ of the differential operator $L_5$ and the inhomogenous polynomial 
$P_{\rm I}(x) $ of (\ref{eq:def3v3}):

\begin{eqnarray}\label{eq:L5}
\fl P_5(x) &=& x^5 \left(1-8x\right)^3\left(1+x\right)^3 \Big(135+3090\,x-629150\,{x}^{2}+6460390\,{x}^{3}-12243595\,{x}^{4} \nonumber \\
\fl && \;-23887460\,{x}^{5}+80746754\,{x}^{6}-237602788\,{x}^{7} +126388752\,{x}^{8}  \nonumber \\
\fl && \;-37648256\,{x}^{9}+49950720\,{x}^{10}-3932160\,{x}^{11}\Big)  \\
\fl P_4(x) &=& 5x^4 \left(1-8x\right)^2\left(1+x\right)^2 \Big( 1512+25377\,x-6996060\,{x}^{2}+106670412\,{x}^{3} \nonumber\\
\fl &&-475126952\,{x}^{4} +96806673\,{x}^{5}+2849916588\,{x}^{6}-5502399670\,{x}^{7} \nonumber\\
\fl &&+9780453960\,{x}^{8}+5163320784\,{x}^{9}-3881744768\,{x}^{10}-715819008\,{x}^{11}\nonumber \\
\fl &&-2127396864\,{x}^{12}+176160768\,{x}^{13}\Big) \nonumber\\
\fl P_3(x) &=&  60x^3 \left(1-8x\right)\left(1+x\right) \Big(2556+26826\,x-11567683\,{x}^{2}+233974966\,{x}^{3} \nonumber\\
\fl &&-1637367768\,{x}^{4}+3360935670\,{x}^{5}+7477700913\,{x}^{6}-32228461430\,{x}^{7} \nonumber\\
\fl &&+ 34977529870\,{x}^{8}-25080039056\,{x}^{9}-116505203984\,{x}^{10}-8874755584\,{x}^{11} \nonumber\\
\fl && +28975368704\,{x}^{12}+19194212352\,{x}^{13}+12468813824\,{x}^{14}-1082130432\,{x}^{15}\Big) \nonumber \\
\fl P_2(x) &=& 60x^2\Big( 23202+90978\,x-100800276\,{x}^{2}+2549503575\,{x}^{3}-24306318922\,{x}^{4}\nonumber\\
\fl && +92736064438\,{x}^{5}-17068101752\,{x}^{6}-721796620433\,{x}^{7}+1283420436692\,{x}^{8} \nonumber\\
\fl && -234713753182\,{x}^{9}-1300729277808\,{x}^{10}+5133789454480\,{x}^{11} \nonumber\\
\fl && +4644548941696\,{x}^{12}-650017517056\,{x}^{13}-1489831145472\,{x}^{14} \nonumber\\
\fl && -997241585664\,{x}^{15}-347227553792\,{x}^{16}+31406948352\,{x}^{17}\Big) \nonumber\\
\fl P_1(x) &=& 120x\Big( 46413+179592\,x-187096284\,{x}^{2}+4466152164\,{x}^{3}-39943265629\,{x}^{4}\nonumber \\
\fl && +141983346692\,{x}^{5}-22051491630\,{x}^{6}-971304694080\,{x}^{7}+1699523699010\,{x}^{8} \nonumber\\
\fl && -484362312312\,{x}^{9}-1466116591440\,{x}^{10}+5048804828832\,{x}^{11} \nonumber\\
\fl && +3936877957248\,{x}^{12}-755770852864\,{x}^{13}-1163160506368\,{x}^{14} \nonumber\\
\fl && -758766698496\,{x}^{15}-242406653952\,{x}^{16}+20937965568\,{x}^{17} \Big) \nonumber\\
\fl P_0(x) &= &7698240+22725360\,x-28148670720\,{x}^{2}+634241959920\,{x}^{3}\nonumber\\
\fl &&-5287126603680\,{x}^{4} +17277689362320\,{x}^{5}-2119806466560\,{x}^{6} \nonumber\\
\fl && -100097290614960\,{x}^{7}+170295740442720\,{x}^{8}-54843096571200\,{x}^{9}\nonumber \\
\fl && -96884451388800\,{x}^{10}+336731369667840\,{x}^{11}+207092810926080\,{x}^{12}\nonumber\\
\fl && -60618583019520\,{x}^{13}-56969296773120\,{x}^{14}-36179450265600\,{x}^{15}\nonumber\\
\fl && -10663262945280\,{x}^{16}+869730877440\,{x}^{17}  \nonumber \\
\fl 
P_{\rm I}(x) &= &7698240+75796560\,x-27539133120\,{x}^{2}+437080734720\,{x}^{3}\nonumber \\
\fl && -2263546745280\,{x}^{4}+2634702988560\,{x}^{5}+9929567400000\,{x}^{6} \nonumber\\
\fl && -19831186297440\,{x}^{7}+17858393541120\,{x}^{8}+2280869253120\,{x}^{9}\nonumber \\
\fl && -8302341242880\,x^{10}+1735690813440\,x^{11}+669914234880\,x^{12}. \label{eq:L5inh}
\end{eqnarray}
 
\noindent
\section{The differential operators $L_3$ and $L_2$ such that $L_5=L_2\oplus L_3$ \label{app:L3L2}}

\begin{eqnarray} \label{eq:L3}
L_3 &\;\;= \;\;&x^3\left( 1+x \right) ^{2} \left( 1-8\,x \right) ^{2}D_x^{\, 3} \; + \nonumber \\ 
& & x^2\left( 1+x \right)  \left( 1-8\,x \right)  \left( 35-158\,x-112\,{x}^{2} \right) D_x^{\, 2} \; +\nonumber \\
&& 6\, x\left( 62-571\,x+687\,{x}^{2}+1616\,{x}^{3}+512\,{x}^{4} \right) D_x \; + \nonumber \\
&& 1188-8460\,x+5712\,{x}^{2}+10752\,{x}^{3}+2304\,{x}^{4}.
\end{eqnarray}

\begin{eqnarray}\label{eq:L2}
\quad\;\; L_2 & \;=\;  &x^2\left( 1+x \right)  \left( 1-8x \right) Q_{2}(x)  D_x^{\, 2} \; +\; 2x\,Q_{1}(x)D_x\;+\;Q_{0}(x). \\ \nonumber\\
Q_{2}(x)&\;=\;&  4+128\,x+816\,{x}^{2}+3455\,{x}^{3}+3386\,{x}^{4}+10331\,{x}^{5}- \nonumber\\
&&19058\,{x}^{6}+36333\,{x}^{7}-30148\,{x}^{8}+23970\,{x}^{9}-4608\,{x}^{10}, \nonumber \\
Q_{1}(x)&\;=\; & 42+1014\,x-584\,{x}^{2}-25649\,{x}^{3}-213659\,{x}^{4}-\nonumber\\ 
&& 288597\,{x}^{5}- 825226\,{x}^{6}+625582\,{x}^{7}-883396\,{x}^{8}- \nonumber\\
&& 148802\,{x}^{9}+221034\,{x}^{10} -621888\,{x}^{11}+129024\,{x}^{12}, \nonumber \\
Q_{0}(x)&\;=\;&396+9216\,x-816\,{x}^{2}-165294\,{x}^{3}-1435806\,{x}^{4}- \nonumber \\
&& 1616220\,{x}^{5}-4745172\,{x}^{6}+3588030\,{x}^{7}-4730706\,{x}^{8}+ \nonumber \\
&& 200916\,{x}^{9}+457740\,{x}^{10}-1540800\,{x}^{11}+294912\,{x}^{12}. \nonumber
\end{eqnarray}

\section{The rational functions of (\ref{eq:S1F21}) \label{app:S1F21}}

\begin{eqnarray}
R_1(x) & =&  -3\frac{(1-8\,x)(1+x)^2\,Q_1(x)}{ x^{10}(1-2\,x)^4} \nonumber\\ \nonumber\\
Q_1(x) &=& 1-1106\,x+6228\,{x}^{2}+360782\,{x}^{3}+574808\,{x}^{4} \nonumber \\
&&750144\,{x}^{5}-909056\,{x}^{6}-444416\,{x}^{7}+24576\,{x}^{8}    \\ \nonumber\\ \nonumber\\
R_2(x) & =& \frac18 \frac{(1-8\,x)(1+x)\,Q_2(x)}{ x^{11}(1-2\,x)^3} \nonumber \\ \nonumber\\
Q_2(x) &=& 1+1072\,x-852\,{x}^{2}+345228\,{x}^{3}-3348324\,{x}^{4} -  20398920\,{x}^{5}- \nonumber\\
&&8922816\,{x}^{6}+40454016\,{x}^{7}+31497216\,{x}^{8}+8126464\,{x}^{9}- \nonumber\\
&&4653056\,{x}^{10}+393216\,{x}^{11} \\ \nonumber\\ \nonumber\\
R_3(x) & =&  \frac18 \frac{(1-8\,x)^2(1+x)^2\,Q_3(x)}{ x^{10}(1-2\,x)^2} \nonumber \\ \nonumber \\ 
Q_3(x) & =& 1+904\,x+5544\,{x}^{2}+254312\,{x}^{3}+423416\,{x}^{4}- \nonumber \\
&& 641856\,{x}^{5}-648704\,{x}^{6}-339968\,{x}^{7}+24576\,{x}^{8} 
\end{eqnarray}

\section*{References}


\begin{thebibliography}{10}

\bibitem{Fisher1984}
Fisher M~E 1984 Walks, walls, wetting, and melting {\em J. Stat. Phys.\/} {\bf
  34} 667--729

\bibitem{Fisher1991}
Fisher M~E, Guttmann A~J and Whittington S~G 1991 Two-dimensional lattice
  vesicles and polygons {\em J. Phys. A: Math. Gen.\/} {\bf 24} 3095--3106

\bibitem{Brak1992}
Brak R, Guttmann A~J and Whittington S~G 1992 A collapse transition in a
  directed walk model {\em J. Phys. A: Math. Gen.\/} {\bf 25} 2437--46

\bibitem{Essam1995}
Essam J~W and Guttmann A~J 1995 Vicious walkers and directed polymer networks
  in general dimension {\em Phys. Rev. E\/} {\bf 52} 5849--5862

\bibitem{Guttmann1993a}
Guttmann A~J and Prellberg T 1993 Staircase polygons, elliptic integrals,
  {H}eun functions, and lattice {G}reen functions {\em Phys. Rev. E\/} {\bf 47}
  R2233--R2236

\bibitem{Essam1993}
Essam J~W 1993 Exact enumeration of parallel walks on directed lattices {\em J.
  Phys. A: Math. Gen.\/} {\bf 26} L863--L869

\bibitem{Gessel1989}
Gessel I and Viennot X~G 1989 Determinants, paths and plane partitions
  {P}reprint

\bibitem{Stembridge1990}
Stembridge J~R 1990 Nonintersecting paths, {P}faffians, and plane partitions
  {\em Adv. Math.\/} {\bf 83} 96--131

\bibitem{Guttmann1998a}
Guttmann A~J, Owczarek A~L and Viennot X~G 1998 Vicious walkers and {Y}oung
  tableaux {I}: Without walls {\em J. Phys. A: Math. Gen.\/} {\bf 31}
  8123--8135

\bibitem{Krattenthaler2000}
Krattenthaler C, Guttmann A~J and Viennot X~G 2000 Vicious walkers, friendly
  walkers and {Y}oung tableaux: {II.} {W}ith a wall {\em J. Phys. A: Math.
  Gen.\/} {\bf 33} 8835--8866

\bibitem{Krattenthaler2003}
Krattenthaler C, Guttmann A~J and Viennot X~G 2003 Vicious walkers, friendly
  walkers, and {Y}oung tableaux. {III.} {B}etween two walls {\em J. Stat.
  Phys.\/} {\bf 110} 1069--1086

\bibitem{Brenti1993}
Brenti F 1993 Determinants of super-schur functions, lattice paths, and dotted
  plane partitions {\em Adv. Math.\/} {\bf 98} 27 -- 64

\bibitem{Guttmann2002a}
Guttmann A~J and V{\"o}ge M 2002 Lattice paths: vicious walkers and friendly
  walkers {\em J. Statist. Plann. Inference\/} {\bf 101} 107--131

\bibitem{Tsuchiya1998}
Tsuchiya T and Katori M 1998 Chiral {P}otts models, friendly walkers and
  directed percolation problem {\em J. Phys. Soc. Japan\/} {\bf 67} 1655--1666

\bibitem{Tabbara2016}
Tabbara R, Owczarek A~L and Rechnitzer A 2016 An exact solution of three
  interacting friendly walks in the bulk {\em J. Phys. A: Math. Th.\/} {\bf 49}
  154004

\bibitem{Brak1997}
Brak R 1997 Osculating lattice paths and alternating sign matrices in {\em
  Proceedings of 9th Formal Power Series and Algebraic Combinatorics
  Conference\/} (Vienna, Austria)

\bibitem{BousquetMelou2006}
Bousquet-M\'elou M 2006 Three osculating walkers {\em J. Phys.: Conf. Ser.\/}
  {\bf 42} 35--46

\bibitem{Brak2013}
Brak R and Galleas W 2013 Constant term solution for an arbitrary number of
  osculating lattice paths {\em Lett. Math. Phys.\/} {\bf 103} 1261--1272

\bibitem{HeunBook}
Ronveaux A, ed. 1995 {\em Heun's differential equation\/} (Oxford: Oxford
  University Press)

\bibitem{Assis2016a}
Assis M, van Hoeij M and Maillard J~M 2016 The perimeter generating functions
  of three-choice, imperfect, and one-punctured staircase polygons {\em J.
  Phys. A: Math. Th.\/} {\bf 49} 214002

\bibitem{Imamoglu2015}
Imamoglu E and van Hoeij M 2015 Maple package {\tt hypergeometricsols}.
  Available at http://math.fsu.edu/\~{}eimamogl/hypergeometricsols/

\bibitem{Imamoglu2016}
Imamoglu E and van Hoeij M 2016 Computing hypergeometric solutions of second
  order linear differential equations using quotients of formal solutions and
  integral bases {\em Preprint submitted to Journal of Symbolic Computation,
  arXiv:1606.01576\/}

\bibitem{Harris1985}
Harris W~A and Sibuya Y 1985 The reciprocals of solutions of linear ordinary
  differential equations {\em Adv. Math.\/} {\bf 58} 119--132

\bibitem{GuessFunc}
Pantone J 2016 {\tt Guess{F}unc}: Automatically forming conjectures about
  differentially algebraic power series. {\em In preparation\/}

\bibitem{Jensen16a}
Jensen I 2016 Square lattice self-avoiding walks and biased differential
  approximants  Submitted to J. Phys. A. Preprint arXiv: 1607.01109

\end{thebibliography}

\end{document}